\documentclass[12pt,preprint]{aastex}

\newcommand{\simless}{\mathbin{\lower 3pt\hbox
      {$\rlap{\raise 5pt\hbox{$\char'074$}}\mathchar"7218$}}} 
\newcommand{\simgreat}{\mathbin{\lower 3pt\hbox
     {$\rlap{\raise 5pt\hbox{$\char'076$}}\mathchar"7218$}}} 

\newcommand{\Msun} {M$_\odot$}
\newcommand{\prodo} {($\Sigma_0 . \kappa_{1.3mm}$)}
\newcommand{\prodm} {($M_D . \kappa_{1.3mm}$)}

\shorttitle{The protoplanetary disks around UX~Ori and CQ~Tau}
\shortauthors{Testi et al.}

\begin{document}


\title{Constraints on properties of the protoplanetary disks around UX~Ori and CQ~Tau}


\author{L. Testi and A. Natta}
\affil{Osservatorio Astrofisico di Arcetri, Largo E.Fermi 5,
       I-50125 Firenze, Italy}
\email{lt@arcetri.astro.it} 

\author{D.S. Shepherd}
\affil{National Radio Astronomy Observatory,P.O. Box O, Socorro, NM 87801}

\and

\author{D.J. Wilner}
\affil{Harvard-Smithsonian Center for Astrophysics, 60 Garden Street,
       Cambridge, MA 02138}




\begin{abstract}
We present Very Large Array observations of the intermediate mass 
pre-main-sequence stars UX Ori and CQ Tau at 7~mm, 3.6~cm, and 6~cm. 
These stars are members of the UX Ori variability class,
where the origin of optical variability is thought to derive from
inhomogeneities in circumstellar disks. 
Both stars are detected at 7~mm but not at longer wavelengths,
which confirms that the millimeter emission is dominated by dust. 

The UX Ori system exhibits a remarkably flat spectral index in the
millimeter range, with $\alpha_{mm}\sim 2$ ($F_{\nu}\propto\nu^{\alpha_{mm}}$).
Two different disk models can reproduce this property: 
i) a physically small disk with optically thick emission, truncated
at a radius about 30 AU, or
ii) a massive ($\sim 0.3-1 M_{\odot}$) disk mainly composed of dust particles 
grown to radii of 10~cm (``pebbles'').
The observations do not spatially resolve the 7~mm emission. 
We discuss implications of these two models and suggest 
observational tests that will discriminate between them.

The CQ~Tau system exhibits a spectral index in the millimeter range of
$\alpha_{mm}\sim 2.6$, consistent with values commonly found for disks 
around pre-main-sequence stars. The observations marginally resolve
the 7~mm emission as an elongated structure with full width at half maximum of 
$2\farcs 4 \times 1\farcs 1$ ($240 \times 110$~AU at 100~pc distance).
The size and inclination of $\sim 63$ degrees (implied by circular symmetry) 
are consistent with flared disk models previously suggested 
to explain the optical colors and polarization properties.
\end{abstract}




\section{Introduction}
\label{sintro}

The study of the circumstellar material around intermediate mass 
pre-main sequence stars has received growing attention in recent years
since it has been realized that most of these systems are surrounded by 
protoplanetary circumstellar disks similar to those found around 
low mass pre-main sequence stars (see e.g. Natta, Grinin \&
Mannings~\citeyear{NGM00}). Herbig~Ae stars (Herbig~\citeyear{H60};
Waters \& Waelkens~\citeyear{WW98})
are particularly interesting because they share the same spectral type
as the much older ``debris'' disk systems, such as $\beta$~Pic
or $\alpha$~Lyrae, of which they may be a preceding evolutionary stage
(Mannings \& Sargent~\citeyear{MS97}; \citeyear{MS00}; Mannings, Koerner \&
Sargent~\citeyear{MKS97}). 

UX~Ori--type stars are intermediate mass pre-main sequence stars,
mostly belonging to the Herbig~Ae class,
that show large and irregular variability at optical wavelengths,
characterized by
sporadic, deep minima  associated with  increasing
polarization. This behavior is generally believed to be 
due to variable line of sight extinction provided by dust clumps 
orbiting about the central star in a flattened structure seen almost edge 
on (Grinin et al.~\citeyear{Gea91}; Grinin~\citeyear{G94}; Natta \&
Whitney~\citeyear{NW00}). The optical spectra of these stars show transient 
absorption features similar to those observed in the ``debris'' disk 
system $\beta$~Pic, and have similarly been interpreted as due to the
infall of evaporating, comet-like bodies in star grazing orbits 
(see the reviews by Lagrange, Backman \& Artymowicz~\citeyear{Lea00} and
Natta et al.~\citeyear{NGM00}).

Single dish and interferometric millimeter wavelength observations 
suggest that UX~Ori--type stars are indeed surrounded by circumstellar
material in a flattened structure rotating about the central star.
The properties of these disks are similar to those
of disks associated with T~Tauri
and other Herbig~Ae stars (Dutrey et al.~\citeyear{Dea96}; Koerner \&
Sargent~\citeyear{KS95}; Mannings \& Sargent~\citeyear{MS97}; \citeyear{MS00}
Natta et al.~\citeyear{Nea97}; \citeyear{Nea00}).
Natta \& Whitney~(\citeyear{NW00}) have shown that scattering of the stellar light
by such disks can easily explain the polarization and colors
typical of UXORs in deep minima, provided that the disk inclination
is in the range $\sim$45$^\circ$ to 70$^\circ$ (Natta \& Whitney~\citeyear{NW00}).
The disks in UX~Ori--type systems are not directly responsible for
the optical variability,
which is caused by a sporadic increase of the optical depth 
along the line of sight, probably due to density inhomogeneities
of matter
within the flared disk atmosphere  (Bertout~\citeyear{B00}).

Recent millimeter interferometric observations
of the class prototype (UX~Ori) revealed that the circumstellar material is
confined within a flattened structure (a circumstellar disk);
the emission is characterized by a flat (F$_\nu\propto\nu^2$) spectral
energy distribution up to 2.7~mm, suggesting that, if the millimeter emission
is entirely due to the thermal emission from the dust grains within the disk,
the disk is either optically thick at millimeter wavelengths or is
composed of very large ($\sim$10~cm) dust particles (Natta et al.~\citeyear{Nea99}).

Similar results have been obtained  for two other  systems in the same
class, WW~Vul and CQ~Tau. In both cases the circumstellar material must
be distributed in a disk as in the case of UX~Ori
(Natta et al.~\citeyear{Nea00}).
It is thus tempting to speculate 
that  UX~Ori--type systems 
are characterized by
relatively evolved  disks,
possibly in the phase of grain growth and planetesimal formation.

In this paper we present the results of new interferometric 
7 millimeter  and centimeter (3.6 and 6~cm) observations of the circumstellar
material surrounding UX~Ori and CQ~Tau. The other UX~Ori--type system
known to have a flat millimeter spectrum (WW~Vul) was excluded from our
sample because of sensitivity considerations (it is the faintest source
of the sample). The aim of these observations
were two-fold: i) to verify whether the flat millimeter wavelength spectral
energy distribution extends up to 7~mm; and ii) to quantify, if any, the
contribution to the millimeter emission due to possible
free-free emission from
an ionized wind. The free-free emission,
if present, will tend to flatten the observed millimeter spectral
energy distribution. The plan of the paper is as follows: in 
Sect.~\ref{sobs} and~\ref{sres} we report the observational details and 
the results; in Sect.~\ref{sdisc} we compare our observations with
detailed disk models; the implications of our results are discussed in
Sect.~\ref{scon}.

\section{Observations}
\label{sobs}

UX~Ori and CQ~Tau were observed with the NRAO\footnote{The National Radio
Astronomy Observatory is a facility of the National Science Foundation
operated under cooperative agreement by Associated Universities, Inc.}
VLA in its most compact (D)
configuration in September and October~2000. The observing parameters 
are reported in Table~\ref{tobs}. Each source was observed in the Q-band 
(43.34~GHz, 0.7~cm), X-band (8.46~GHz, 3.6~cm) and C-band (4.86~GHz, 6~cm),
the latter observations were obtained at the same time as the Q-band
observations using all the antennas not equipped with 43~GHz receivers
(6 for UX~Ori and 5 for CQ~Tau). At high frequency we used a fast-switching
observing cycle with total duration ranging from one to three minutes depending
on the distance between the source and the chosen complex gain calibrator
and on the elevation of the sources. At X- and C-band we used a standard 
observing sequence with $\sim$5~min cycles between source and calibrator.
Hourly pointing at X-band on the calibrators were used to correct for 
pointing drifts during the Q-band observations. The flux density scale was set
by observing 3C286 and/or 3C48, the calibration is expected to be accurate 
within 10-20\% depending on the waveband, with the less accurate being the
higher frequency band.
All data editing and calibration has been performed using standard tasks 
within the NRAO AIPS software package. Self-calibration was not possible 
due to the faintness of the sources. All images presented here have 
been produced using the AIPS IMAGR task with natural weighting of the 
($u,v$) datasets. No correction for primary beam attenuation has been applied,
unless differently noted.
At 6~cm, the image rms is higher than theoretical expectation by a
factors of about 2 (UX~Ori) and 5 (CQ~Tau) due to the very sparse 
$(u,v)$ coverage and the presence of strong confusing sources in the
field of view far off the phase center.




\section{Results}
\label{sres}


\subsection{7~mm emission}

\begin{figure}
\plotone{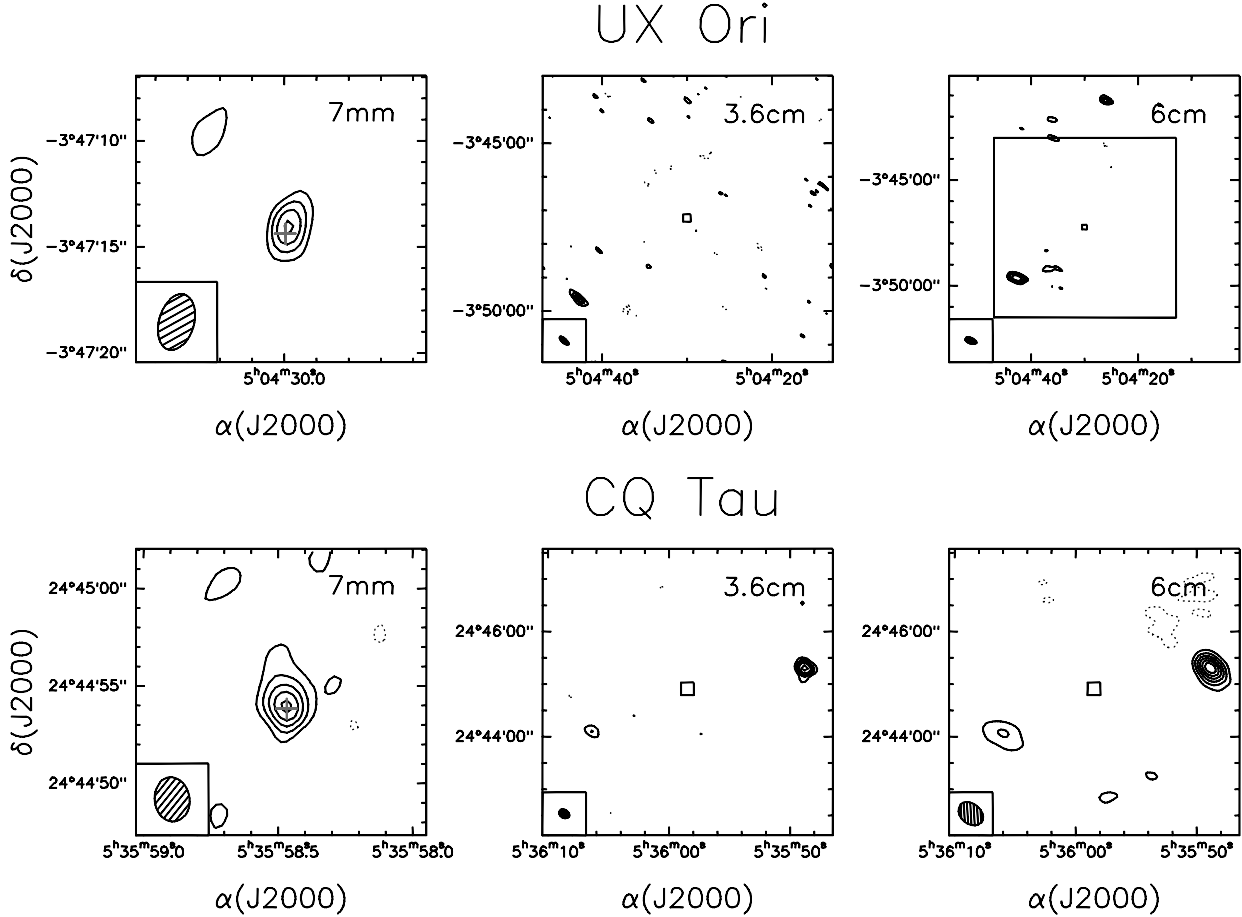}
\caption{VLA radio continuum maps centered on UX~Ori (top) and CQ~Tau (bottom).
For each source we show the 7~mm map on the left and the 3.6~cm  and 6~cm maps
on the
right. The crosses on the left panels mark the optical positions of the target
stars corrected for proper motion following the Hipparcos measurements
(Perryman et al.~\citeyear{Hipp}). The small square at the center of
the 3.6~cm map shows the
area covered by the 7~mm panel on the left. The large square in the UX~Ori 6~cm map
shows the area covered by the 3.6~cm map.
Contour levels for UX~Ori are: --0.3, 0.3 to 0.9 by 0.2~mJy/beam at 7~mm,
--0.09, 0.09 to 0.6 by 0.015~mJy/beam and 1.1~mJy/beam at 3.6~cm, and
--0.2, 0.2 to 0.6 by 0.15~mJy/beam at 6~cm; for CQ~Tau: --0.45, 0.45 to 1.65
by 0.3~mJy/beam at 7~mm, --0.09, 0.09 to 0.6 by 0.015~mJy/beam and
1.1~mJy/beam at 3.6~cm, and --0.5, 0.5 to 3.7 by 0.4~mJy/beam at 6~cm.
Negative contours are shown as dotted lines.
\label{fcontmaps}}
\end{figure}

Both sources were clearly
detected at 7~mm. 
In Figure~\ref{fcontmaps} we
show the 7~mm contour plots of the region surrounding the target stars.
In Table~\ref{tres} we report the properties of the 7~mm emission 
along with  stellar parameters from the
literature. For each source we report distance and spectral type
of the target star from Natta et al.~(\citeyear{Nea00}, \citeyear{Nea99}),
stellar age and mass 
as derived from the theoretical pre-main sequence evolutionary tracks
of Palla \& Stahler~(\citeyear{PS99}), then the VLA peak position,
integrated flux, and deconvolved size (or upper limit) as obtained by fitting
a gaussian to the 7~mm source. 
In both cases the peak of the 7~mm emission is coincident (within our 
positional uncertainties of $\sim 0\farcs3$) with the position of the optical star
corrected for proper motion
as derived from the Hipparcos catalogue
(Perryman et al.~\citeyear{Hipp}). 

The millimeter emission is
unresolved in UX~Ori.
In the case of  CQ~Tau, the 7~mm emission appears to be 
marginally resolved and elongated in the north-south direction (see
Table~\ref{tres}). We find consistent results, albeit with lower 
signal-to-noise, by fitting a gaussian on a slightly higher resolution image of CQ~Tau
obtained  using the ``robust'' weighting scheme available with the 
AIPS IMAGR task.
Assuming a circularly symmetric, geometrically thin disk,
the measured inclination of the disk axis from the line of sight is
63$^{+10/-15}$~deg,
in excellent agreement with that predicted by the Natta \&
Whitney~(\citeyear{NW00}) models, who derive a value of 66~deg to
explain the polarization of light during the deep optical emission minima.
It is interesting to note that, based on CO(2--1) interferometric observations,
Mannings \& Sargent~(\citeyear{MS00}) derive an upper limit to the size of the
gaseous disk smaller than our 7~mm continuum observation, this difference may be
due to a different surface brightness sensitivity.
Clearly higher resolution observations of both the gaseous and dust components
are needed to confirm the exact disk geometry.

\subsection{3.6 and 6~cm emission}

There is no detection of either  source
at 3.6 and 6~cm.  In the last two columns of
Table~\ref{tres}
the 3.6 and 6~cm upper limits (3$\sigma$) at the 7~mm sources positions
are given.
There are, however, a few sources
far from the stellar positions (see Fig.~\ref{fcontmaps}). 
For the sake of completeness, in Table~\ref{tother} we report positions
and fluxes of these sources, as well as the NVSS counterpart name and
integrated flux (when found in the NVSS catalogue, Condon et
al.~\citeyear{Cea98}). Following Fomalont et al.~(\citeyear{Fea91}) 
and Windhorst et al.~(\citeyear{Wea93}) we can estimate the number of 
expected background extragalactic radio sources in our fields: at 6~cm
over a 200 square arcmin area we expect $\sim$3 sources brighter than 0.5~mJy,
while at 3.6~cm over 50 square arcmin, $\sim$2.5 sources brighter than 0.1~mJy are
expected. These predictions are in rough agreement with the number of
centimeter sources we observed, and suggest that they may be extragalactic,
which would also be consistent with the non-thermal spectral index
of three of them (see
Table~\ref{tother}). In any case, none of these cm wavelength radio sources
is directly related to our target stars and their circumstellar material and
we will not discuss them further.

\subsection{The millimeter spectra of UX~Ori and CQ~Tau}
\label{srsed}

To obtain the millimeter spectrum of the circumstellar material
surrounding our target stars, we combined our measurements with 
available millimeter interferometric and single dish continuum observations 
from Natta et al.~(\citeyear{Nea99}) and Mannings \&
Sargent~(\citeyear{MS97}; \citeyear{MS00}). We  include both
interferometric and single dish measurements and assume
that the mm-wave emitting region is confined to a compact region 
around the star. This assumption is supported by the fact that the 1.3~mm
single dish and interferometer fluxes are consistent with each other
(see also the discussion in Natta et al.~\citeyear{Nea99}).
The cm-wavelength upper limits can be used to constrain 
the possible contribution of
free-free emission to the flux measured at 7~mm.  Following
Mannings \& Sargent~(\citeyear{MS97}) and using the 3.6~cm upper limits reported
in Table~\ref{tres}, the contribution to the 7~mm flux from an ionized stellar
wind, with a spectral index 0.6 (Panagia \& Felli~\citeyear{PF75}), would be 
0.24 and 0.2~mJy for UX~Ori and CQ~Tau, respectively. This estimate
of the upper limit for the free-free emission confirms
the earlier conclusion that the millimeter wavelength emission in these
systems is due to thermal dust emission with a very small (if any) contribution
from ionized gas (Mannings \& Sargent~\citeyear{MS97}; Natta et al.~\citeyear{Nea99}).

\begin{figure}
\plotone{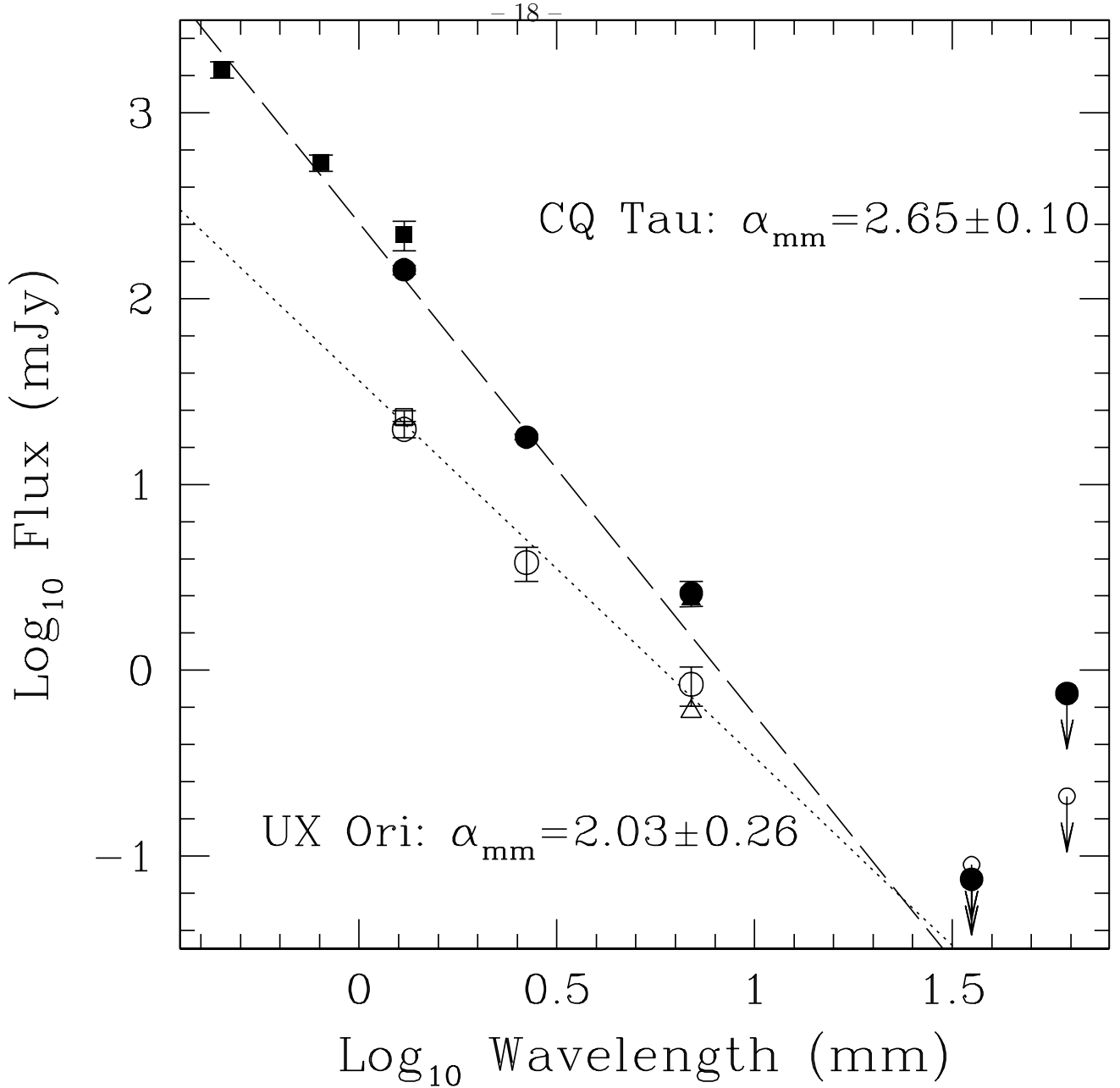}
\caption{Millimeter wavelength spectral energy distributions for 
UX~Ori (open symbols) and CQ~Tau (filled symbols) circumstellar disks.
The 1.3 and 2.7~mm interferometric fluxes and all the single
dish measurements 
are from Natta et al.~(\citeyear{Nea99}) and Mannings \& Sargent~(\citeyear{MS97};
\citeyear{MS00}).
The lines show the results of power-law fits of the form
F$_\nu\propto\nu^{\alpha_{mm}}$ for UX~Ori (dotted) and CQ~Tau (dashed),
the resulting values of $\alpha_{mm}$ are also shown.
At 3.6 and 6~cm we show the 3$\sigma$ upper limits from this work.
The open triangle at 7~mm shows the minimum dust thermal emission flux for
UX~Ori after subtracting the maximum possible free-free contribution (see text).
The filled triangle is the similarly corrected point for CQ~Tau.
Single dish measurements are shown as squares, interferometric ones as circles.
\label{fmmspec}}
\end{figure}

In Figure~\ref{fmmspec} we show the results of power law fits
(F$_\nu\propto\nu^{\alpha_{mm}}$) to the millimeter spectral energy 
distributions.
The values of $\alpha_{mm}$ that we derive are
rather low.
The value $\alpha_{mm}=2.03\pm0.26$ in UX~Ori, in particular,
is among the lowest of those
found for Herbig~Ae/Be stars (Natta et al.~\citeyear{NGM00}) and is 
significantly smaller than the average values found for T Tauri stars
(Dutrey et al.~\citeyear{Dea96}). 
In Figure~\ref{fmmspec} we also show the 7~mm flux 
of UX~Ori with the maximum correction for free-free emission as
calculated above. The point, shown as an open triangle, would still be
consistent with the thermal dust emission fit we derive.
For the CQ~Tau system, we obtain $\alpha_{mm}=2.65\pm0.10$. In this case
the maximum possible contribution from free-free emission to the 7~mm flux
is so small ($<8\%$) that it is barely visible in the Figure.

These values of $\alpha_{mm}$ are not consistent with optically thin
thermal emission from standard interstellar dust grains ($\alpha_{mm}\sim 4$;
Draine \& Lee~\citeyear{DL84}). This result is not uncommon for circumstellar
disks around young stars and can be understood if  the emission is
optically thick or, alternatively,
if  grains in  circumstellar disks have grown to sizes
significantly larger than those of grains in
the   diffuse interstellar medium 
(Beckwith \& Sargent~\citeyear{BS91}; Pollack et al.~\citeyear{Pea94};
Natta et al.~\citeyear{Nea99}).
In fact, for grains with physical size larger than the observing wavelength,
the emissivity is just  the geometrical cross section, and a 
flat spectral index at millimeter wavelengths may be an indication of the 
presence of centimeter-size dust particles (``pebbles'').
We will discuss these possibilities
in detail for our target systems in the following section.



\section{Comparison with disk models}
\label{sdisc}

The spectral energy distributions (SEDs) of Herbig Ae stars
in general, and of UX~Ori and CQ~Tau in particular,
can be explained reasonably well by disk models
which take into account
the vertical temperature gradient caused by the stellar
radiation impinging on the disk surface
(Natta et al.~\citeyear{Nea00}; Chiang et al.~\citeyear{Cea00}). 
In these models, the dust features,
such as the 10 $\mu$m emission feature, and the mid-infrared
continuum originate in the hotter optically thin surface layers
of the disk (the disk atmosphere), while the emission
at longer wavelengths is due to dust in the cooler disk
midplane.   Using the simple two-layer approximation developed
by Chiang \& Goldreich~(\citeyear{CG97}) Natta et al.~(\citeyear{Nea99},
\citeyear{Nea00}) found that Herbig Ae disks are in hydrostatic
equilibrium (flared disks; Kenyon
\& Hartmann~\citeyear{KH87}). The grains in the disk atmosphere are
a mixture of amorphous silicates and some other material, 
possibly carbonaceous grains or metallic iron, which dominate the
opacity in the UV. These atmospheric grains cannot be much larger
than $\sim$1 $\mu$m. In the disk midplane grains are probably much larger.

The properties of the grains in the disk midplane can only be
studied at millimeter wavelengths, where the emission in
many objects becomes optically thin. Its dependence on
wavelength reflects the way dust opacity depends on $\lambda$,
and may tell us about the grain size.
However, optical depth effects may still be important even at
1.3~mm. As discussed in Natta et al.~(\citeyear{Nea99}) for UX~Ori,
both an optically thin disk where grains have grown to very
large size (``pebbles") and a small disk, optically thick in the
millimeter,
have spectral index $\sim 2$ and may fit equally well
the observed SED at all wavelengths. As we will see in the
following,
the new 7~mm data presented in this paper allow us 
to  clarify   the issue, even if they are still not
sufficient for deciding it.

To this purpose, we have computed a
large grid of models for both UX~Ori and CQ~Tau
 following Chiang \& Goldreich~(\citeyear{CG97})
and Natta et al.~(\citeyear{Nea00}). For each model, we fix the stellar parameters
as in Table~\ref{tres}, and take the values of the inclination
to the line of sight of 60$^\circ$ and 66$^\circ$ for UX~Ori and CQ~Tau
respectively, as derived by Natta \& Whitney~(\citeyear{NW00}).
We vary the disk outer radius $R_D$ and
the surface density profile, defined as:
\begin{equation}
\Sigma = \Sigma_0 \> \Big({{R}\over{R_0}}\Big)^{-p}
\end{equation}
where $R_0$ is a reference radius. Both $p$ and $\Sigma_0$ are varied over
a wide range of values. The disk mass $M_D$ is uniquely determined  once 
$\Sigma_0$, $p$ and $R_D$ are fixed. Since we are interested
in the disk emission at long wavelengths, the results do not depend
on the disk inner radius.

We describe the  opacity
of the dust in the disk midplane as a power-law function
of wavelength with exponent $\beta$:
\begin{equation}
\kappa_\nu = \kappa_{1.3mm}\> \Big({{\lambda}\over{1.3mm}}\Big)^{-\beta}
\end{equation}
The long-wavelength flux depends on the properties of dust in the
atmosphere  only if and when the disk midplane is optically thin (to its
own radiation). 
We choose dust properties for the grains in the disk atmosphere
as in Natta et al.~(\citeyear{Nea00}); any other choice that fits the
SEDs at shorter wavelengths would not affect our conclusions
significantly.
These ``toy'' models are very simplified and can be questioned on a variety of 
grounds (see e.g. Dullemond~\citeyear{D00}). However, they are very useful
for the general understanding of the implications of our millimetric 
observations.

\subsection{UX~Ori}
\label{sduxori}

\begin{figure}
\plotone{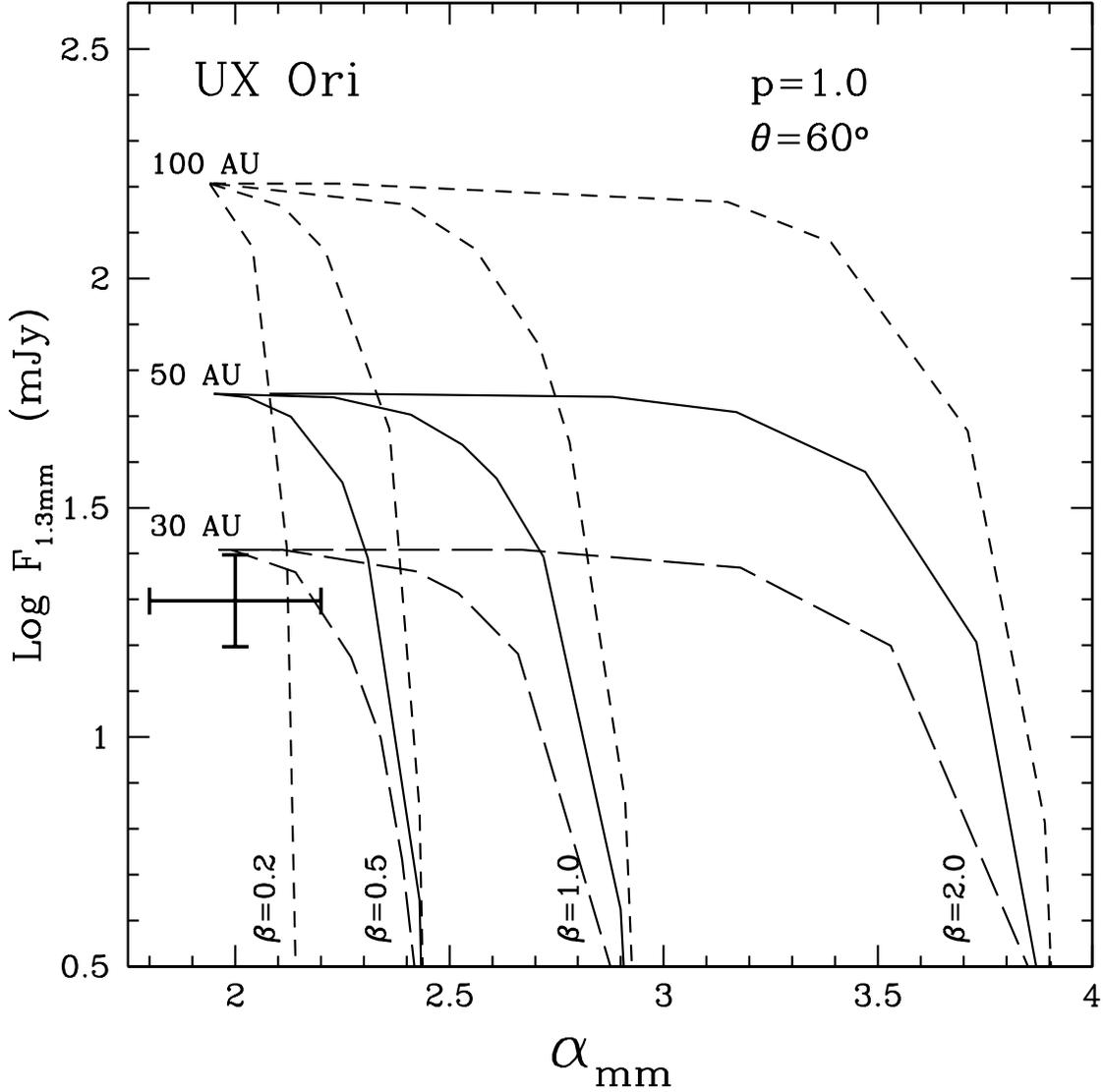}
\caption{This figure shows the values of the 1.3mm flux as function of the
spectral slope $\alpha_{mm}$ (between 1.3 and 7~mm)
predicted by disk models of UX~Ori.
Each curve is computed for disks having fixed values of the outer radius
$R_D$ and of the dust opacity slope $\beta$, as labeled, and decreasing
values of the product $\Sigma_0 \times \kappa_{1.3mm}$.
In all models,
the surface density varies with radius as $R^{-1}$. The stellar parameters
are T$_\star$=8600 K, L$_\star$=42 L$_\odot$, M$_\star$=2.3 \Msun,
D=450 pc.  The inclination angle is $\theta=60^o$ (Natta \& Whitney 2001).
The cross shows the observed  position of UX~Ori in this diagram.
\label{uxori}
}
\end{figure}

\begin{figure}
\plotone{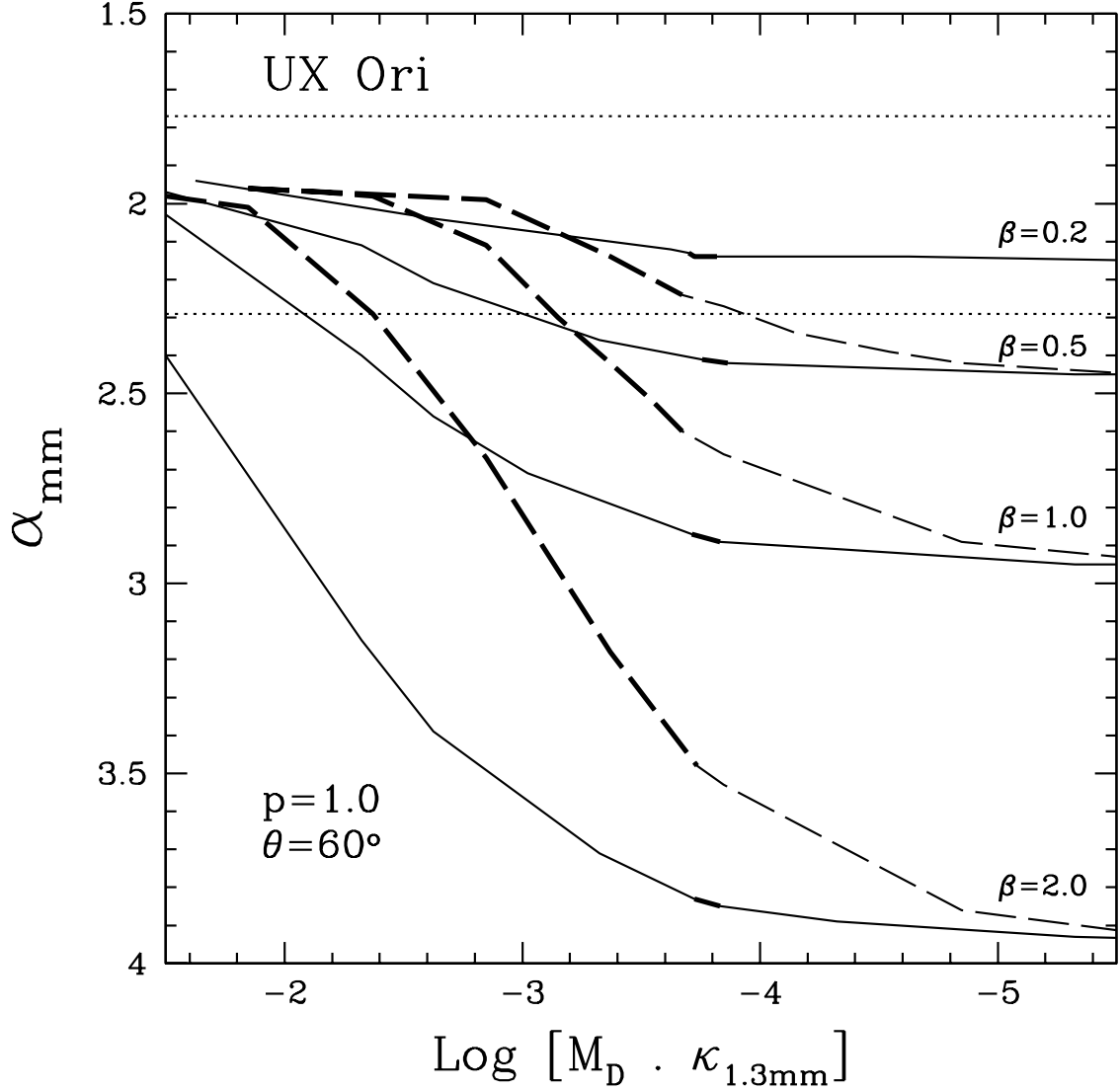}
\caption{
The figure shows  $\alpha_{mm}$ as function of the product \prodm,
where $M_D$ is the disk mass (dust only) in units of \Msun\
and $\kappa_{1.3mm}$ the dust opacity in cm$^2$ g$^{-1}$ of dust.
The models are the same shown in Fig.~\ref{uxori} with $R_D=100$ AU
(solid lines) and $R_D=30$ AU (dashed lines). The thicker part
of each curve shows the region where the model-predicted
$F_{1.3mm}$ is equal to the observed value of $19.8 \pm 2$ mJy
(Natta et al. 1999).
The  two horizontal dotted lines enclose the observed 
$\alpha_{mm}=2.06 \pm 0.26$ spectral index.
\label{mass}
}
\end{figure}

The  results for UX~Ori are shown in Fig.~\ref{uxori}, which plots
the computed flux at 1.3mm as function of the spectral index
$\alpha_{mm}$ between 1.3 and 7~mm.
Each curve is computed varying
\prodo\ for fixed values of $R_D$ and $\beta$, as
labeled. \prodo\ decreases from the upper left to the lower right of
the diagram for all models.
The models shown in the Figure have been computed
for $p=1.0$, but the results are not significantly different for
$p=0.5$ and $p=1.5$.
All curves ($F_{1.3mm}$, $\alpha_{mm}$) have a similar shape.
For large values of \prodo, the disks are optically
thick at  mm wavelengths, $F_{1.3mm}$ is maximum and $\alpha_{mm}\sim 2$.
As \prodo\ decreases, first the emission at 7~mm becomes
optically thin: $F_{1.3mm}$ retains its optically thick value but
$\alpha_{mm}$ increases. If \prodo\ decreases further,  the
emission at 1.3~mm becomes optically thin, $F_{1.3mm}$ decreases
and $\alpha_{mm}$ tends to its optically thin value of $2+\beta$.
The maximum value of $F_{1.3mm}$ depends on the inclination $\theta$
and on $R_D$ (see also Natta et al.~\citeyear{Nea99}).
For fixed values of $p$ and $R_D$, the disk mass is proportional
to $\Sigma_0$. Fig.~\ref{mass} 
plots the values of $\alpha_{mm}$ as function of the quantity
\prodm\ for $R_D=100$ AU (solid lines) and $R_D=30$ AU (dashed lines).

As shown by the cross in Fig.~\ref{uxori}, UX~Ori
has a low
1.3~mm flux ($F_{1.3mm}=19.8\pm 2$ mJy; Natta et al. 1999)
and a very flat spectral index.
There are  two families of models that can account for
both properties at the same time: small disks ($R_D \simless$ 30 AU)
which are still optically thick  at 7~mm or
disks with very large grains, such that
$\beta\simless 0.2$. 
We have computed, using Mie theory, the  cross section
at millimeter wavelengths for large silicates and ices,
based on  data for a variety of minerals from the Jena database 
and for the astronomical silicates of Draine \& Lee~(\citeyear{DL84}).
In all cases we find that  grains with $\beta\sim 0.2$
between 1.3 and 7~mm  have radii of about 10~cm; their  opacity
is very low, of order of $2\times 10^{-2}$ cm$^2$ g$^{-1}$ (this is the
{\it dust} only mass opacity, assuming that the grains have a density
of 3.5~g\,cm$^{-3}$; no correction has been applied for the gas to 
dust mass ratio). 
This is about 3 times lower than the value in Natta et al.~(\citeyear{Nea99}),
based on Fig.~4 of Miyake \& Nakagawa~(\citeyear{MN93}), who considered
grains of silicates and ices with density $\sim$1.2~g\,cm$^{-3}$.
For these values of the opacity, the observed fluxes can only be
reproduced by disks with a very massive dust component of about 
$0.0025- 0.01$~\Msun,
with little dependence on
 other disk parameters such as $p$ and $R_D$. This dust mass 
is a factor of three to ten 
higher than the largest dust masses measured in Herbig Ae/Be 
systems (Natta et al.~\citeyear{NGM00}). When corrected for a gas to dust ratio 
of 100 by mass, the corresponding {\it total} disk mass would reach 
0.3--1~\Msun.

Disks that are still optically thick at 7~mm up to their outer edge,
where most of the 7~mm flux is emitted, have $\alpha_{mm}\sim 2$.
In order to fit the low millimeter flux of UX~Ori, they
need to be ``truncated'' at $R_D\simless 30$ AU, as shown in Fig.~\ref{uxori}. 
The dust mass required depends on dust properties; it is  at least
$1.3 \times 10^{-3}$~\Msun\ for 
grains typical of  pre-main sequence disks ($\kappa_{1.3mm}\sim 1$
cm$^2$ g$^{-1}$, $\beta\sim 1.0$, see also Fig.~\ref{mass}). 
Again, assuming a gas to dust ratio of 100 by mass, the total disk mass
would be 0.13~\Msun.
This is a rather large value, but not unique among Herbig Ae/Be systems.

\begin{figure}
\plotone{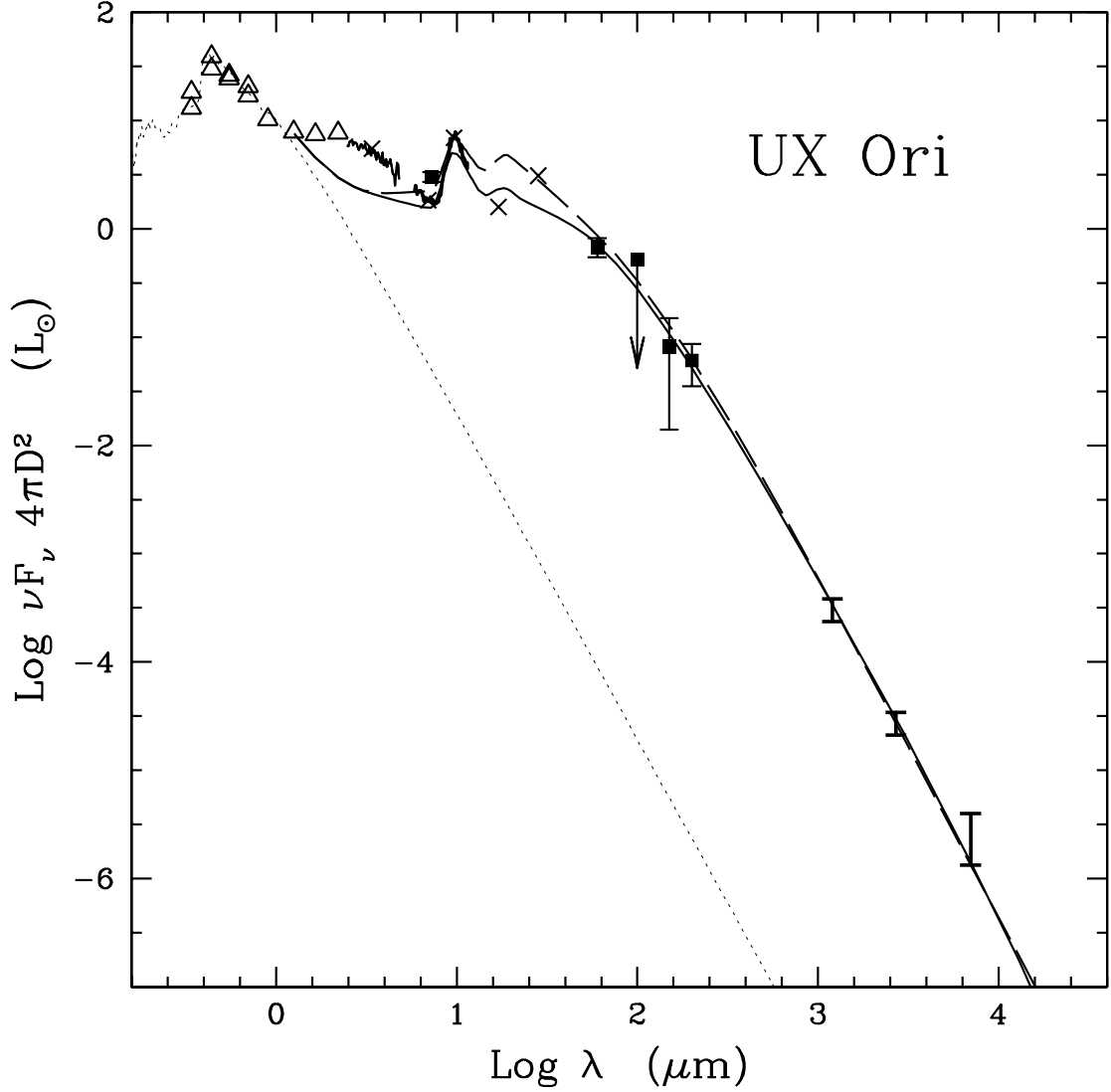}
\caption{Spectral Energy distribution  of UX~Ori corrected
for an extinction $A_V$=0.5 mag (see Natta et al.~\citeyear{Nea99} for
references).
The interferometric fluxes at 1.3, 2.7 and 7~mm are shown by the
error bars.
The dotted line shows the stellar emission. The solid  line plots the
predictions of the small ($R_D$=28 AU), optically thick 
($M_D=1.3 \times 10^{-3}$ \Msun of dust,
$\kappa_{1.3mm}=1$ cm$^2$ g$^{-1}$) disk discussed in the text.
The other disk parameters are $\beta=1$, $p=1$.
The dashed line shows the prediction of the ``pebble" disk models
($\beta$=0.2, $\kappa_{1.3mm}=2\times 10^{-2}$ cm$^2$ g$^{-1}$ of dust)
with $M_D=1$ \Msun. Also in this case $p=1$.
\label{sed}
}
\end{figure}

Both kind of disks, ``pebble'' and ``truncated'', fit well
the UX~Ori SED over the whole range of observed
wavelengths, as shown in Fig.~\ref{sed}.
However, $R_D\sim$ 30 AU is a definite lower limit for the ``truncated'' disks
radius, since smaller disks would not emit enough flux in the 
mid-infrared. A 30 AU disk can still scatter enough stellar light
to reproduce 
the polarization and colors of
the light during photometric minima
(Natta \& Whitney~\citeyear{NW00}).
In a pebble disk, both the properties of the star at minima
and the shape of the mid-infrared emission require the
presence of rather small grains on the disk surface.
This is consistent 
with the idea that grains  grow to large size first in the
disk midplane; however, the
condition $\alpha_{mm}\sim 2$ constrains
the mass of dust in small grains to be
$\simless$ 1\% of  the mass of pebble grains
(Natta et al.~\citeyear{Nea99}). Note, finally, that the condition that
the disk is flared requires that the total disk mass is dominated by the
gaseous component.


\subsection{CQ~Tau}
\label{sdcqtau}

\begin{figure}
\plotone{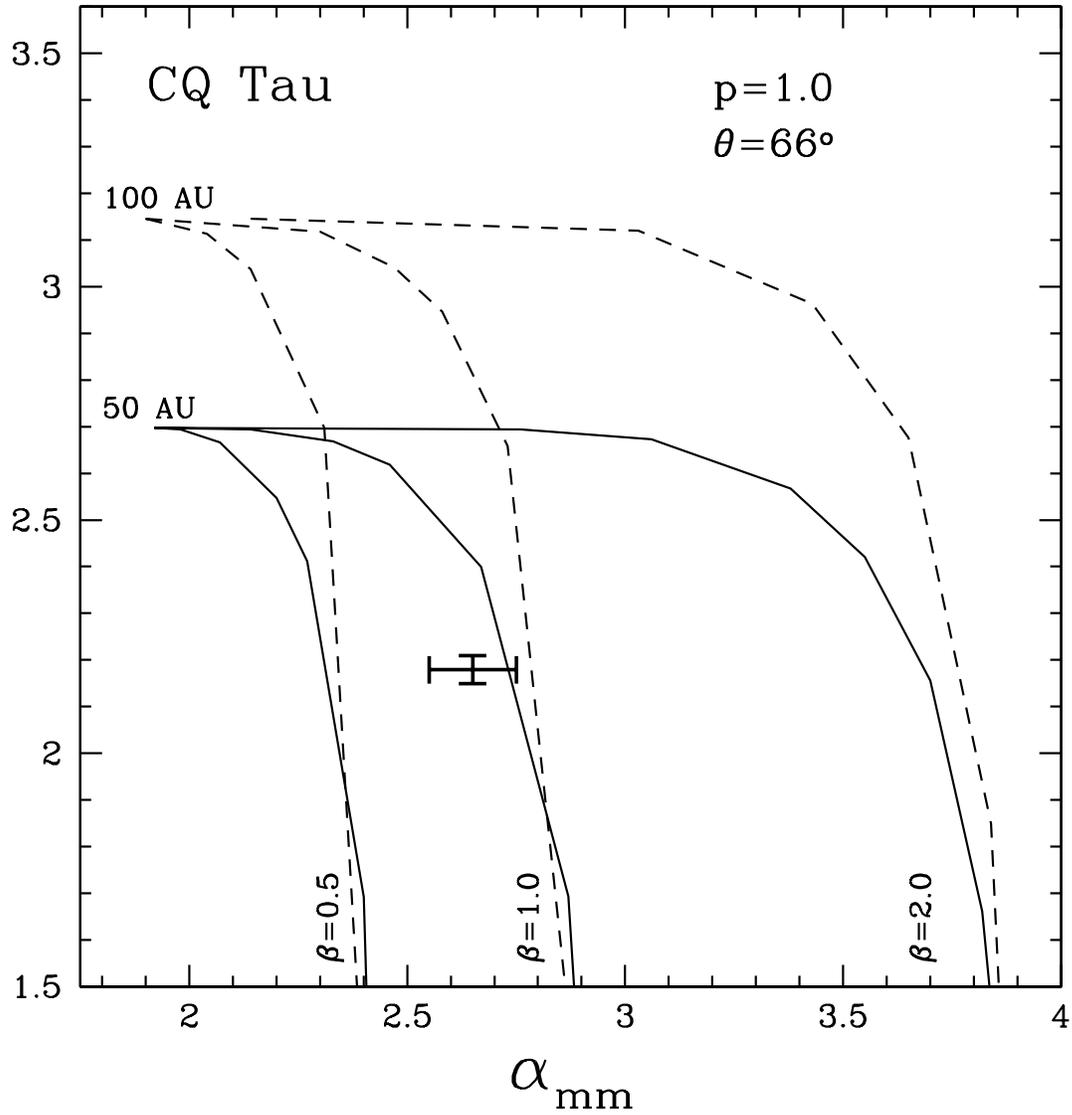}
\caption{Same as Fig.3 for CQ Tau.  The stellar parameters are 
T$_\star$=7500 K, L$_\star$=5 L$_\odot$, M$_\star$=1.5 \Msun,
D=100 pc. The inclination angle is $\theta=66^o$ (Natta \&
Whitney~\citeyear{NW00}).
\label{cqtau}
}
\end{figure}

The CQ~Tau system is closer to the Sun and the central star is cooler than 
UX~Ori. The millimeter wavelength spectral energy distribution is not as flat 
as in UX~Ori, and the value of $\alpha_{mm}$ we derive is only 
marginally smaller than the typical value found in T~Tauri systems.
The location of CQ~Tau on
the ($F_{1.3mm}$, $\alpha_{mm}$) diagram is shown in Fig.~\ref{cqtau}.
It is consistent with a disk having $R_D\sim$ 100 AU, as observed,
and grains with $\beta\sim 1.0$, typical of pre-main sequence disks
(Beckwith \& Sargent 1991).
The mass of dust in the  disk  is $\sim 1.5 \times 10^{-4}$~\Msun, 
again typical
of Herbig Ae  and T Tauri stars (Natta et al. 2000).



\section{Discussion and Conclusions}
\label{scon}

UX~Ori is certainly an intriguing system. The circumstellar disk
may be very small and optically thick, in which case it must be
``truncated'' at R$\simless 30$~AU. How this could happen is 
unclear. It may be an evolutionary effect, but the system is not
a member of a dense stellar cluster (cf. Testi et al.~\citeyear{TPN99})
and the expected
timescale required for the star to photoevaporate the outer regions of the disk
is too long to explain such a sharp truncation at the
estimated system age of $\sim$3~Myrs (Hollenbach et al.~\citeyear{Hea00};
Hollenbach priv. comm.).
A more plausible explanation for such a small  disk is the 
presence of a nearby companion.  To create a
tidal truncation of 30~AU, the orbital separation should be roughly
60--100~AU (Lubow \& Artymowicz~\citeyear{LA00}). No companion 
was found in the moderate (0.4\arcsec\ or 200~AU) resolution search by 
Pirzkal et al.~(\citeyear{Pea97}), although Hipparcos data  indirectly
suggests the presence of a close companion with
a minimum separation of about 20 mas ($\sim 10$ AU at distance of 450 pc;
Bertout et al.~\citeyear{Bea99}).

Alternatively, the UX~Ori disk may be  more ``normal" in size and have
grown very large (10~cm) particles in its midplane. Such a disk must be
very massive ($\sim$0.25--1 \Msun\ for a gas-to-dust mass
ratio of 100 or $M_D/M_\star \sim 0.1-0.4$), much more massive than
any known disk around Herbig Ae and T Tauri stars and very close to the limit
for gravitational stability (Shu et al. 1990).
Note that this mass estimate assumes a gas-to-dust ratio of 100
and an opacity for the ``pebble'' dust as derived in Sect.~\ref{sduxori}.
A moderate gas depletion may lower this estimate of the
{\it present day} disk mass. However, 
the {\it dust} mass in UX~Ori, up to 0.01~\Msun\ for ``pebble'' opacity,
would be exceptionally high, much higher than the dust mass
derived from the analysis of millimeter fluxes in other
pre-main sequence stars (Natta et al.~\citeyear{NGM00}); moreover the
{\it initial} disk mass before gas depletion would still be very high.

With the available information, the choice between  a truncated
disk and a larger, pebble disk
can only rely 
on plausibility arguments. Pebble disks must be very massive.
However,  uncertainties on the 
opacity and on the exact disk structure
make it difficult to determine the mass of the disk accurately.
Two possible observational tests may be able to discriminate
between the possibilities: i) to search for a possible companion, and
ii) to resolve the disk emission at millimeter wavelengths.
A 0.2\arcsec\ separation
binary, even with a relatively large luminosity difference,
can easily be resolved with modern near-infrared high resolution techniques,
unless the geometry is very unfavorable. 
The detection of a close companion  which could
truncate the disk around the primary star would support this hypothesis.
 A more direct approach would be
to resolve the disk and determine $R_D$. As discussed earlier, a large
disk radius would  require pebble grains, and could be resolved at 7~mm with the VLA,
or at higher frequencies with ALMA.
A large disk radius would not be consistent with ``normal" disk grains.

Is UX~Ori a unique object? If this were the case, albeit interesting
{\it per se}, it would not shed light on the structure and evolution
of typical protoplanetary disks. However, it
appears more likely
that the properties of the disk around UX~Ori are shared by a class of 
systems which may represent a well defined disk evolutionary stage.
As discussed in Sect.~\ref{sintro}, WW~Vul, which also shares a similar
optical photometric variability, has a millimeter spectrum similar 
to UX~Ori with $\alpha_{mm}=2.2\pm 0.4$ between 1.3 and 2.7~mm \cite{Nea00}.
A set of promising
candidates with flat {\it sub}--millimeter SEDs has been recently discussed
by Meeus et al.~(\citeyear{Mea01}). Millimeter wavelength observations of
these systems are needed to verify whether the disk properties are similar 
to that of UX~Ori.

CQ~Tau appears to have a disk which is similar to other pre-main sequence
disks. The primary advantage of observing CQ~Tau is
 the fact that it is near-by
(D $\sim 100$~pc). Our observations with VLA in its most compact (D)
configuration marginally resolve its
7~mm emission.  They
are consistent with the disk inclination predicted by Natta \&
Whitney~(\citeyear{NW00}), thus providing an indirect support for their
models. 
The sensitivity and  resolution of the VLA in more extended
configurations at 7~mm 
offer the possibility to further resolve the disk emission and 
directly probe the density and temperature structure of the emitting dust
within the circumstellar disk of a pre-main sequence star. Very few
young disk systems have been recently resolved in the millimeter continuum:
the binary system L1551 (Rodr\'{\i}guez et al.~\citeyear{Rea98}), and the
TTauri systems HL~Tau and TW~Hyd (Wilner \& Lay~\citeyear{WL00}; Wilner et
al.~\citeyear{Wea00}).

Our observation do not offer any clues about the possible link 
between the UX~Ori--type optical variability and disk properties.
We can only confirm the earlier suggestion that the disk surrounding 
CQ~Tau is seen at the inclination inferred by fitting its optical light
polarization properties with flared disk models. Since a prediction
of these models is that the star is observed through the outer edges of
the disk atmosphere, a plausible guess is that the sporadic variability 
may be caused by disk atmosphere inhomogeneities (Bertout~\citeyear{B00}; 
Natta \& Whitney~\citeyear{NW00}).

\acknowledgments
We thank Barry Clarke and the VLA-AOC staff for 
for their help while planning and executing the observations and
for nice scheduling at the VLA. Support from CNR--ASI grants ARS--99--15 and
1/R/27/00 to the Osservatorio Astrofisico di Arcetri is gratefully
acknowledged. DJW acknowledges support from 
NASA Origins of Solar Systems Program grant NAG5-8195.
This research has made use of the SIMBAD database, operated at CDS,
Strasbourg, France.




\clearpage

\begin{deluxetable}{rcccccc}
\tabletypesize{\scriptsize}
\tablecaption{VLA observing parameters. \label{tobs}}
\tablewidth{0pt}
\tablehead{
\colhead{Parameter} & \colhead{0.7~cm}   & \colhead{3.6~cm}   & \colhead{6~cm}
& \colhead{0.7~cm}   & \colhead{3.6~cm}   & \colhead{6~cm}
}
\startdata
&\multicolumn{3}{c}{\bf UX~Ori}&\multicolumn{3}{c}{\bf CQ~Tau}\\
\tableline
Phase center $\alpha$(J2000) &\multicolumn{3}{c}{05$^h$04$^m$29.99$^s$}
             & \multicolumn{3}{c}{05$^h$35$^m$58.47$^s$}\\
$\delta$(J2000) & \multicolumn{3}{c}{$-$03$^\circ$47$^\prime$14.3$^{\prime\prime}$}
             & \multicolumn{3}{c}{$+$24$^\circ$44$^\prime$54.1$^{\prime\prime}$}\\
Observing date  & 10Sep00 & 05Sep00 & 10Sep00 & 02Oct00 & 02Oct00 & 05Oct00 \\
Configuration & D & D& D& D & D& D\\
Duration$^a$ (hrs)& 8 & 1 & 8& 7 & 1 & 7\\
Frequency (GHz) & 43.34&8.46 &4.86& 43.34&8.46 &4.86\\
Bandwidth (MHz) & 4$\times$50 & 4$\times$50 & 4$\times$50& 4$\times$50 & 4$\times$50 & 4$\times$50\\
Gain calibrators fluxes 0501$-$019 (Jy) & 0.83 & 1.05 & 1.18&--&--&--\\
    0559$+$238 (Jy) &--&--&--& 0.48 & 0.51 & 0.52\\
Primary beam  &1\arcmin &5.$\!^\prime$3 &9.$\!^\prime$3&1\arcmin &5.$\!^\prime$3 &9.$\!^\prime$3\\
Largest structure that can be imaged&43\arcsec &3\arcmin & 5\arcmin&43\arcsec &3\arcmin & 5\arcmin\\
Synthesized beam (FWHM) & $2\farcs 7\times 1\farcs 6$ & 
                   $20\arcsec\times 8\arcsec$ &
                   $33\arcsec\times 17\arcsec$&
                   $2\farcs 3\times 1\farcs 8$&
                   $12\arcsec\times 9\arcsec$ &
                   $31\arcsec\times 23\arcsec$\\
Position Angle &$-$15$^\circ$ & 49$^\circ$ & 69$^\circ$&7$^\circ$ &58$^\circ$& 49$^\circ$ \\
Noise (mJy/beam) &0.1 &0.03 & 0.07& 0.15 & 0.025 & 0.25\\
 \enddata
\tablenotetext{a}{This is the total duration of the observations, calibration
overhead at 0.7~cm was in the range 40--50\%}


%

\end{deluxetable}
\clearpage

\begin{deluxetable}{lrcrrcccccll}
\tabletypesize{\scriptsize}
\tablecaption{Source properties. \label{tres}}
\tablewidth{0pt}
\tablehead{
\colhead{Star} & \colhead{D$^a$}  & \colhead{Sp.T.$^a$}&\colhead{Age$^b$}
& \colhead{M$_\star^b$}& \colhead{$\alpha_{7mm}$}& \colhead{$\delta_{7mm}$}&
\colhead{F$_{7mm}$}&\colhead{FWHM}&\colhead{p.a.}&\colhead{F$_{3.6cm}$}&
\colhead{F$_{6cm}$}\\
&\colhead{(pc)}&&\colhead{(Myr)}
&\colhead{(M$_\odot$)}&\colhead{(J2000)}&\colhead{(J2000)}&
\colhead{(mJy)}&\colhead{(AU)}&\colhead{(deg)}&\colhead{(mJy)}&
\colhead{(mJy)}
}
\startdata
UX~Ori&450&A3&2.2&2.3&5:04:29.98&$-$3:47:14.1&0.84$\pm$0.2&$<$700&--&$<$0.09&$<$
0.2\\
CQ~Tau&100&F2&9.0&1.5&5:35:58.47&$+$24:44:54.2&2.6$\pm$0.4&240$\pm$30$\times$110
$\pm$30&2$\pm$13&$<$0.075&$<$0.75\\
\enddata
\tablenotetext{a}{Natta et al.~(\citeyear{Nea99}; \citeyear{Nea00});}
\tablenotetext{b}{Ages and masses have been computed by comparing the stellar parameters
given by Natta et al.~(\citeyear{Nea99}; \citeyear{Nea00}) with the pre-main
sequence evolutionary tracks of Palla \& Stahler~(\citeyear{PS99}).}
\end{deluxetable}

\clearpage

\begin{deluxetable}{rccccrc}
\tabletypesize{\scriptsize}
\tablecaption{Parameters of the sources detected at centimeter wavelengths. \label{tother}}
\tablewidth{0pt}
\tablehead{
\colhead{Identifier} & \colhead{$\alpha$} & \colhead{$\delta$} &
\colhead{$F_{3.6~cm}$}   & \colhead{$F_{6~cm}$} & \colhead{NVSS}
& \colhead{$F_{20~cm}$}\\
& \colhead{(J2000)} & \colhead{(J2000)} &
\colhead{(mJy)}   & \colhead{(mJy)} & \colhead{Counterpart} & \colhead{(mJy)}
}
\startdata
J0504259$-$034115 & 05:04:25.9 & $-$03:41:15 & --  & 2.2 & NVSSJ0504257$-$034112 & 4.7\\
J0504427$-$034938 & 05:04:42.7 & $-$03:49:38 & 3.9 & 3.1 & -- & --\\
J0535488$+$244519 & 05:35:48.8 & $+$24:45:19 & 2.5 & 3.1 & NVSSJ0535489$+$244517 & 6.8\\
J0536064$+$244406 & 05:36:06.4 & $+$24:44:06 & 0.3 & 1.4 & -- & --\\
\enddata
\tablecomments{The identifiers are derived from the J2000 coordinates 
of the sources. The integrated fluxes have been corrected for the primary
beam attenuation, since the sources are far away from the phase center the
correction is large (especially at 3.6~cm) and the fluxes are affected 
by large uncertainties.}

\end{deluxetable}

\clearpage


\clearpage 

\end{document}